\documentclass[twocolumn,preprintnumbers,amsmath,amssymb,superscriptaddress,email]{revtex4}

\usepackage[pdftex]{graphicx}
\usepackage{dcolumn}
\usepackage{bm}
\usepackage[latin1]{inputenc}
\usepackage{amsmath}

\newcommand{\ud}[1]{{#1^{\dagger}}}

\newcommand{\mean}[1]{\langle#1\rangle}

\begin{document}


\title{Comment on ``Nonlinear photoluminescence spectra from a quantum dot-cavity system revisited'' by Yao \emph{et al.} and ``Photoluminescence from Microcavities Strongly Coupled to Single Quantum Dots'' by Ridolfo \emph{et al.}}

\author{F. P.~Laussy}
\affiliation{School of Physics and Astronomy. University of Southampton, Southampton, SO171BJ, United Kingdom.}
\author{E. del Valle}
\affiliation{School of Physics and Astronomy. University of Southampton, Southampton, SO171BJ, United Kingdom.}
\author{C. Tejedor}
\affiliation{F\'{\i}sica Te\'orica de la Materia Condensada, Universidad Aut\'onoma de Madrid, 28049, Spain.}

\date{\today}

\begin{abstract}
  We refute the criticisms of our work on strong-coupling in the
  presence of an incoherent pumping.
\end{abstract}

\maketitle

Our description of strong-coupling in
semiconductors~\cite{laussy08a,laussy09a,delvalle09a} has been
recently criticized by two groups: Ridolfo \emph{et
  al.}~\cite{arXiv_ridolfo09a} and Yao \emph{et
  al.}~\cite{arXiv_yao09a}.

Both groups of authors make the same statements: they claim that our
master equation is flawed, on the ground that its domain of
convergence is bounded, they both propose to use exclusively a thermal
bath for the reservoir of excitations of the cavity instead of our
most general case (Yao \emph{et al.} require a thermal bath also for
the excitonic reservoir whereas Ridolfo \emph{et al.} allow
independent pumping and decay coefficients for the Quantum Dot (QD))
and they all claim a better agreement than our model with experimental
data.

We show in this comment that their criticisms are erroneous and that
the alternatives they propose to fulfill them are well known particular
cases that, in their approximations, are also erroneous.

Both groups focus on our boson model only, and in fact only on our
Letter on the topic~\cite{laussy08a}. Many of their statements are
already addressed in our full boson text~\cite{laussy09a} and more
crucially in its fermion counterpart~\cite{delvalle09a}. Throughout,
we made clear that the boson model is adequate either in the limit of
small pumping, or in the limit of bosonic~0D system. The interest of
the boson case is in its analytical solutions, that allow to explain
transparently fundamental features of pumping, such as the effect of
the effective quantum state on the spectral shapes. We will show that
the analytical solutions for the fermion models from
\cite{arXiv_ridolfo09a} and \cite{arXiv_yao09a} are in fact valid only
in trivial cases (namely, the boson limit or the uncoupled limit).

In the following, we shall clarify, on the one hand, that our approach
does not suffer from any inconsistency or pathology, and on the other
hand, that in the particular cases where the reservoirs of excitations
are thermal baths, our model also applies (by enforcing the ad-hoc
constrain in the equation) and show the errors made by the
approximations of Refs.~\cite{arXiv_ridolfo09a,arXiv_yao09a}.

\section{Validity of our master equation}

Our model describes the linear regime (of vanishing excitations)
analytically and the nonlinear regime semi-analytically. Its master
equation reads:
\begin{subequations}
  \label{eq:SunNov9134217GMT2008}
  \begin{eqnarray}
    \frac{d\rho}{dt}&=&i[\rho,H_a+H_\sigma+H_{a\sigma}]\\
    &+&\frac{\gamma_a}2(2a\rho\ud{a}-\ud{a}a\rho-\rho\ud{a}a)\label{eq:MonOct5021932BST2009}\\
    &+&\frac{P_a}2(2\ud{a}\rho a-a\ud{a}\rho-\rho a\ud{a})\label{eq:MonOct5021954BST2009}\\
    &+&\frac{\gamma_\sigma}2(2\sigma\rho\ud{\sigma}-\ud{\sigma}\sigma\rho-\rho\ud{\sigma}\sigma)  \\
    &+&\frac{P_\sigma}2(2\ud{\sigma}\rho \sigma-\sigma\ud{\sigma}\rho-\rho\sigma\ud{\sigma})\,.
  \end{eqnarray}
\end{subequations}
where $a$ is the cavity mode and $\sigma$ the exciton in the QD.  The
cavity is always a bosonic mode. In Ref.~\cite{laussy08a}
and~\cite{laussy09a}, $\sigma$ is also a Bose operator, that we note
$b$ for clarity, while in Ref.~\cite{delvalle09a}, $\sigma$ is a
fermion operator, describing a two-level system~\footnote{We have been
  aware of and considering the Fermi dynamics from the start, as one
  can see from the initial version arXiv:0711.1894v1 of our
  manuscript~\cite{laussy08a} although it eventually retained only the
  boson results, postponing the general but non-analytical case to
  Ref.~\cite{delvalle09a}. All our statements in the linear and boson
  models, even in November 2007, were confronted with the fermion case
  for validity, in contrast with our critics, as is shown in this
  comment.}. This choice of~$\sigma$ as a Bose operator addresses two
important cases: $i$) the limit of vanishing pumpings (even if the QD
is indeed a fermion emitter) and $ii$) the case where the excitons
follow bose statistics. The latter case could be realized in large
QDs, that recover the physics of quantum wells where excitons are
known to behave as good bosons~\cite{kavokin_book07a}.

The main reproach of~\cite{arXiv_ridolfo09a,arXiv_yao09a} is that in
Eq.~(\ref{eq:SunNov9134217GMT2008}), the effective pumping
rates~$P_{a,\sigma}$ can vary independently of the effective decay
rates $\gamma_{a,\sigma}$~\footnote{They both also make the reproach
  that~$\sigma$ is limited to Bose statistics but this is because they
  are unaware---or do not want to consider---our main line of work,
  for which we can only point them to Ref.~\cite{delvalle09a}.}. They
observe that this can lead to some divergences beyond some critical
values of pumping, as we have ourselves discussed before, and they
conclude that the model is flawed. They propose instead to use thermal
reservoirs, that do not exhibit such divergences, at all values of
pumping.

A thermal reservoir for a bosonic mode $a$ (with frequency $\omega_a$
and $H_a=\omega_a\ud{a}a$) at temperature $T$ leads to the effective
rate of excitation:
\begin{equation}
  \label{eq:MonNov3195527GMT2008}
  P_a=\kappa_a \bar n_T\,.
\end{equation}
with~$\bar n_T$ given by the reservoir Bose-Einstein distribution. It
vanishes at $T=0$. In thermal equilibrium, the system is loosing
excitations at a larger rate of:
\begin{equation}
  \label{eq:MonNov3195553GMT2008}
  \gamma_a=\kappa_a(1+\bar n_T)=\kappa_a+P_a\,.
\end{equation}
The parameter $\kappa_a$ is the \emph{spontaneous emission} (SE) rate
at $T=0$.  The steady state thermal equilibrium reads
\begin{equation}
  \label{eq:FriOct31164539GMT2008}
  n_a=\mean{\ud{a}a}=\frac{P_a}{\gamma_a-P_a}=\frac{P_a}{\kappa_a}=\bar n_T\,.
\end{equation}
At very high temperatures, as the effective income of particles
approaches the outcome, $P_a\approx\gamma_a$, the number of particles
remains finite, since $P_a<\gamma_a$. As long as $\gamma_a\neq0$, any
combination of parameters $\gamma_a,P_a$ corresponds to a given
thermal bath (with $\kappa_a=\gamma_a-P_a$ and $T>0$).

The linewidth of the optical spectrum of emission is:
\begin{equation}
  \label{eq:SunNov16124512GMT2008}
  \Gamma_a\equiv\gamma_a-P_a=\kappa_a\,.
\end{equation}
and is independent of temperature (i.e., of the population of the
mode), since it is always equal to the spontaneous emission decay rate
$\kappa_a$.

It is clear from the above results, that a bosonic thermal bath cannot
provide gain and does not exhibit any line-narrowing in its
luminescence spectrum. A thermal bath is a medium of \emph{loss} as
$P_a<\gamma_a$ by definition. 

A thermal reservoir is, however, a particular case. In
out-of-equilibrium conditions, especially under externally applied
pumping, one can expect deviations from the thermal paradigm.

In contrast to the thermal case, a \emph{gain} medium can be derived
with bosonic baths out-of-equilibrium. This is discussed in textbooks,
e.g., in Chapter~7 of Gardiner and Zoller's
text~\cite{gardiner_book00a}. A linear gain can be obtained with an
``inverted'' harmonic oscillator maintained at a negative temperature
$-T'$. The effect in the master equation is that the effective
parameters are now given by a relation opposite to
Eqs.~(\ref{eq:MonNov3195527GMT2008}-\ref{eq:MonNov3195553GMT2008}):
\begin{equation}
  \label{eq:FriOct2145057GMT2009}
  \gamma_a=G_a\bar m_{-T'}\,, \quad  P_a=G_a (1+\bar m_{-T'})=G_a+\gamma_a \,.
\end{equation}
In this way, $G_a$ is the gain or input of particles into the mode at
zero temperature. Obviously, given that now $P_a>\gamma_a$, there is
no stationary solution to the master or rate equations. Since the
absence of this stationary solution is the source of confusion in
Refs~\cite{arXiv_ridolfo09a,arXiv_yao09a}, we quote here Gardiner and
Zoller's comment, p°216:

\begin{quote}
  \emph{If $\gamma < \kappa$} [that is $\gamma_a<P_a$ with our
  notation]\emph{, there is no stationary situation, and the amplifier
    gives a signal that increases without limit. Essentially, the
    power being fed into the cavity cannot escape fast enough. (Of
    course the idea of an inverted medium which maintains its
    inversion independent of power output is not exactly valid, and
    depletion effects will then need to be considered. The system is
    then essentially a laser).}
\end{quote}

Obviously, an ever-growing population will always be stopped by some
external physical effect (the sample will have burnt, the reservoir
will be depleted, etc\dots). There is however nothing pathologic in
this behaviour. In particular, this does not invalidate the results
for values below the critical pumping rates. In the case
where~$\gamma_c>P_c$ ($c=a,b$), there is a physical solution to
the dynamical master equation, starting at~$t=0$: at all
(\emph{finite}) times, there is a valid master equation, with positive
trace, normed to unity, etc\dots\ However, indeed, the system diverges
with time. There is no unphysical behaviour or flaw of some sort.  Not
all dynamical systems have a steady state, some because they are
oscillatory, others because they increase without bounds.

In our work~\cite{laussy08a, laussy09a, delvalle09a}, we have
naturally considered the configurations which admit a steady state. We
have even given analytical solutions for their domain of convergence,
supported by a clear physical picture~\footnote{The limit of validity
  of the boson master equation is given by a total outcome of
  excitation larger than the income $\gamma_a+\gamma_b>P_a+P_b$, that
  is, $\Gamma_a+\Gamma_b>0$ (in the notations of~\cite{laussy09a}) If
  one of the oscillators has a net gain ($\Gamma_b<0$), there is an
  extra condition for convergence that has a clear physical motive:
  not only mode~$a$ is constrained to be neatly lossy, $\Gamma_a>0$,
  but also the income through the ``inverted'' mode $b$, given by
  $P_b$, must be smaller than its total outcome, which is given by
  $\gamma_b$ plus the effective (Purcell) decay through cavity
  emission. That is, $P_b< \gamma_b + \frac{4
    (g^\mathrm{eff})^2}{\Gamma_a}$ where
  $g^\mathrm{eff}=g/\sqrt{1+[2\Delta/(\Gamma_a+\Gamma_b)]^2}$, and
  $=g$ at resonance.}.

In a microcavity QED system (a QD in a microcavity), which is a
complicated solid state system, open to many sources of
excitations~\cite{gies07a,arXiv_winger09a}, one should also include gain effects in the most
general case. Granted all together, the microscopic coefficients are
likely terms of the form:
\begin{subequations}
  \label{eq:FriOct2151738GMT2009}
  \begin{eqnarray}
    &\gamma_a=\kappa_a (1+\bar n_T)+G_a\bar m_{-T'}\,,\\
    &P_a=\kappa_a\bar n_T+G_a (1+\bar m_{-T'})\,,
  \end{eqnarray}
\end{subequations}
that is, including loss media and gain media. Net losses in the case
of a cavity mode comes, among other reasons, from the fact that the
photons can escape the cavity through the imperfect mirrors. Net gain
could come from surrounding off-resonance or weakly coupled QDs, high
energy QD levels or the wetting layer. A gain medium can be obtained
by the very configuration realized with self-assembled QDs in a
microcavity. Gardiner and Zoller, p°140 (ibid) study a bath of
uncorrelated two-level emitters that are kept in average in the
excited state. Quoting them again:

\begin{quote}
  \emph{Note that there is no restriction on $N^+_a$ and $N^-_a$}
  [that is, in their notations, the number of emitters in the excited
  and ground states, respectively, that provide the pump and decay
  terms for the bosonic mode]\emph{---this means that $N^+_a>N^-_a$ is
    permissible. Physically such a population inversion could only be
    maintained by some kind of pumping, as indeed happens in a laser.}
\end{quote}
and as indeed could happen in an incoherently pumped
microcavity.  The previous discussion on gain-media that applies for
the cavity mode can also be extended for bosonic emitters (QDs, in our
case). In this case, for example, the electron-hole pairs that form
excitons inside the QD decay from the wetting layer (at higher energy)
by, e.g., emitting phonons with the corresponding energy
difference. Such phonons will not be reabsorbed to bring back the
electron-hole pair to the wetting layer, leading to a net source of
particles. Net losses take place through the spontaneous decay into
leaky modes.

In general, there is thus no reason to restrict the master
equation~(\ref{eq:SunNov9134217GMT2008}) designed to account
effectively for the largest possible amount of physical effects, to a
given type of reservoirs (namely, thermal baths). Our study aims at
the greatest level of applicability and generality and provides the
tools for the understanding of any case. Therefore, better than trying
to apply at the theoretical level any criteria (other than
convergence) to choose $\gamma_a$, $P_a$, we prefer to consider them
independent. So far, the dynamics of
lines~(\ref{eq:MonOct5021932BST2009}-\ref{eq:MonOct5021954BST2009})
found its most important domain of applicability with atom lasers and
polariton lasers~\cite{holland96a,imamoglu96b,porras03a,rubo03a,
  laussy04c,schwendimann06a,schwendimann08a,doan08a}, that is, systems
where a condensate (or coherent state) is formed by scattering of
bosons into the final state from another state rather than by
emission. In both cases, scattering or emission, the process is
stimulated. In this case the income and outcome of particles is a
complicated function of the distribution of excitons (or polaritons)
in the higher $k$-states. See, e.g., Ref~\cite{imamoglu96b}. In a
pulsed experiment, it is typically time dependent (see, e.g., Fig.~2
of Ref.~\cite{laussy04b}). Line narrowing is a natural feature of this
dynamics in such systems.

It remains, of course, possible that a particular experiment
corresponds to a thermal bath of excitation. In this case, a fitting
analysis with our model should indicate this constrain in correlations
between the $\gamma$ and~$P$ coefficients. As a result of this
analysis, one will then understand that the given system refers to the
particular cases of Eqs.~(\ref{eq:MonNov3195527GMT2008})
and~(\ref{eq:MonNov3195553GMT2008}).

The above discussion concerns the bosonic mode. We now turn to the
fermion mode.

The thermal equilibrium case, at temperature $T$, gives the
counterpart of the boson case (given above):
\begin{subequations}
  \label{eq:MonOct5000945BST2009}
  \begin{align}
    \gamma_\sigma&=\kappa_\sigma (1+ \bar n_T)=\kappa_\sigma+P_\sigma\,,\\
    P_\sigma&=\kappa_\sigma \bar n_T\,,
  \end{align}
\end{subequations}
where $\kappa_\sigma$ is the Einstein $A$-coefficient and $P_\sigma$
is the Einstein $B$-coefficient. The steady state is the
\emph{Fermi-Dirac distribution}:
\begin{equation}
  \label{eq:FriOct31175956GMT20082}
  n_\sigma=\frac{P_\sigma}{\gamma_\sigma+P_\sigma}=\frac{P_\sigma}{\kappa_\sigma+2P_\sigma}=\frac{\bar n_T}{2\bar n_T+1}=\Big(e^{\frac{\omega_\sigma}{k_BT}}+1\Big)^{-1}\,.
\end{equation}
The maximum occupation for the emitter is~$1/2$, at infinite
temperature. It is, therefore, not possible to invert the two-level
system with a thermal bath, (where, again, $\gamma_\sigma>P_\sigma$),
which is a well known result. In the fermion case, even a gain-medium
does not lead to a divergence thanks to the intrinsic saturation of a
two-level system. The emission spectrum is also a Lorentzian, with
effective linewidth:
\begin{equation}
  \label{eq:SunNov16125821GMT2008}
  \Gamma_\sigma\equiv \gamma_\sigma+P_\sigma=\kappa_\sigma+2P_\sigma
\end{equation}
which broadens from the decay rate at zero temperature,
$\kappa_\sigma$ when the temperature increases. 

This elementary discussion illustrates the notorious fact that thermal
reservoirs are unable to achieve population inversion of a two-level
system~\cite{briegel93a}. Therefore, the choices of excitation
reservoirs of Yao \emph{et al.}~\cite{arXiv_yao09a}, forbids lasing in
such systems, which is already contradicted by
experiments~\cite{arXiv_nomura09a}.

The issue of achieving gain with a master equation for a fermion
emitter has been extensively discussed in the lasing literature, in
particular in the one-atom laser
case~\cite{agarwal90a,cirac91a,mu92a,horak95a,briegel96a,loffler97a,meyer98a,benson99a,koganov00a,florescu04a,karlovich08a,li09a},
and it is typically described theoretically also by considering an
effective negative-temperature thermal bath. A popular notation to
consider gain and dissipation is~\cite{briegel96a}:
\begin{equation}
  \label{eq:FriOct2101847GMT2009}
  P_\sigma=\Gamma_\sigma s \,, \quad \gamma_\sigma=\Gamma_\sigma (1-s)\,.
\end{equation}
With $\Gamma_\sigma>0$ (the linewidth broadening) and $0\leq s \leq
1$, including both the situations with net losses ($s<1/2$) and gain
($s>1/2$). In this case:
\begin{equation}
  \label{eq:FriOct31175956GMT200822}
  n_\sigma=\frac{P_\sigma}{\gamma_\sigma+P_\sigma}=s\,.
\end{equation}
This parameterization is not linked to any particular experimental
realization but is designed to separate the physical effects that
lead to line broadening, $\Gamma_\sigma$, from those that change the
population, $s$. Apart from that, it is equivalent to consider
directly the effect of varying the effective decay and pumping
parameters, $\gamma_\sigma$ and $P_\sigma$, as we have done in
Refs~\cite{delvalle09a} and \cite{arXiv_gonzaleztudela09a}.

The Jaynes-Cummings model couples the fermionic and bosonic modes. It
has been extensively studied in the case of thermal cavity bath and
some gain-medium for the emitter (which is also the case of
Ref.~\cite{arXiv_ridolfo09a}). We studied it in its most general form,
using the master equation~(\ref{eq:SunNov9134217GMT2008}). Again, if a
particular constrain arises from a given realization of the reservoirs
of excitations, such as
(\ref{eq:MonNov3195527GMT2008}-\ref{eq:MonNov3195553GMT2008}),
(\ref{eq:FriOct2145057GMT2009}), (\ref{eq:FriOct2151738GMT2009}),
(\ref{eq:MonOct5000945BST2009}) or some other case, this would appear
in our unconstrained case from correlations following these trends.
We expect, as we previously discussed, that a successful statistical
analysis would indeed inform about underlying microscopic details of
the excitation scheme.

\section{Validity of the proposed substitutes to our work}

In the case of Ridolfo \emph{et al.}~\cite{arXiv_ridolfo09a}, only the
photonic mode was excited thermally while the excitonic pump was still
allowing an unconstrained pumping and decay. Yao \emph{et
  al.}~\cite{arXiv_yao09a}, on the other hand, require both modes to
be excited by thermal baths.

Thermal baths in the linear (boson) model reduce to results identical
to the spontaneous emission of an initial state that is a mixture of
excitons and photons in the ratio of
population~$n_a(t=0)/n_b(t=0)=P_a/P_b$. The thermal character of the
bath merely prevents renormalization of the linewidths and of the Rabi
frequencies. The ratio $P_a/P_b$ still determines the possibility to
resolve the line-splitting. This fundamental consequence of the
effective quantum state is independent of any choice of the
reservoirs.  It is a general result that we have amply discussed
before~\cite{laussy08a,laussy09a,delvalle09a} and that is
``rediscovered'' by Ridolfo \emph{et al.}

Ridolfo \emph{et al.}~\cite{arXiv_ridolfo09a} otherwise have mistakes
in their formulas, that certainly bias their analysis. For instance,
their parameters $n_a$ and $C$ should read, in their notations:
(cf.~their Eq.~(7)~\&~(8))
\begin{subequations}
  \label{eq:MonOct5005157BST2009}
  \begin{eqnarray}
    n_a&=&\frac{P_a}{\gamma_a}+\frac{g^2}{\gamma_a}\frac{(\gamma_a+\gamma_x)(\gamma_a P_x - \gamma_x  P_a)}{g^2(\gamma_a+\gamma_x)^2 +\gamma_a \gamma_x |\tilde\omega_a^* -\tilde\omega_x|^2}\,,\\
    C&=&\frac{g}{\tilde\omega_a^* - \tilde\omega_x}(n_a - n_x)\,.
  \end{eqnarray}
\end{subequations}
Also, their spectra are normalized to~$\sqrt{2\pi}n_a$, again
apparently as an error since they compare them directly to ours which
are normalized to unity. These mistakes do not seem to be a misprint,
given that the authors state in their paper:
\begin{quote}
  \emph{Although at low pump intensities, our approach and that of
    Ref. [Laussy et al.] essentially represent models of a linear
    Bose-like dynamics of two coupled harmonic oscillators, nontrivial
    differences can be appreciated.}
\end{quote}

There should be no difference in the limit of vanishing pump. In fact,
once corrected as above, their formulas and lineshapes do converge to
our results at low pumps.

\begin{figure}[tbhp]
  \centering
  \includegraphics[width=\linewidth]{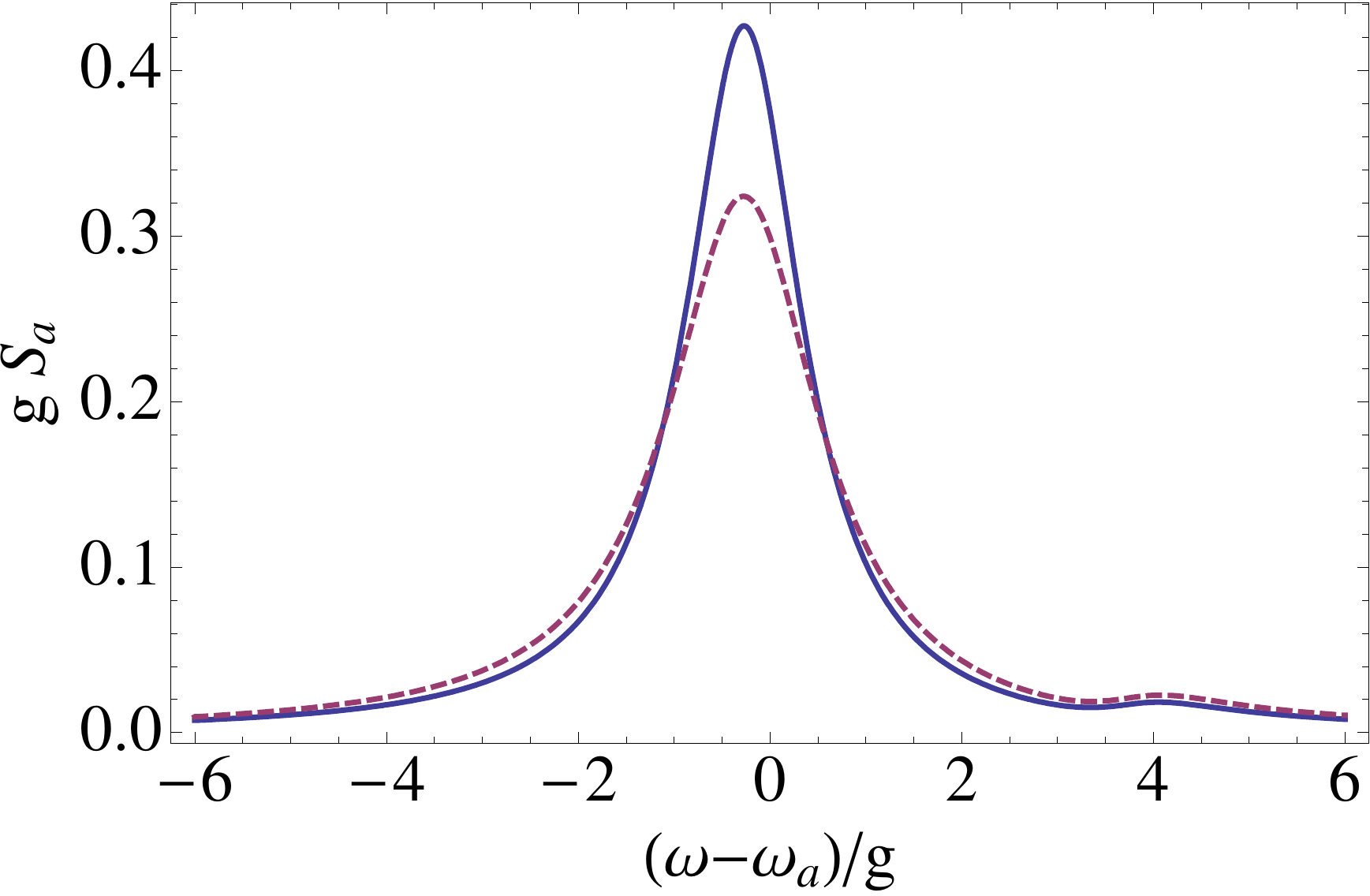}
  \caption{Cavity emission for the parameters of Fig.~1(a) of Ridolfo
    \emph{et al.}~\cite{arXiv_ridolfo09a}: $\Delta=-3.6g$,
    $\gamma_\sigma=1.48g$, $P_a=0.49g$, $P_\sigma=0.0078g$. Like in
    their figure---but with the correct
    formulas~(\ref{eq:MonOct5005157BST2009})---we compare: in
    solid-blue, the case where $\gamma_a=1.96g$ and in dashed-purple
    $\kappa_a=1.96g$ ($\gamma_a=\kappa_a+P_a$, thermal bath). Both
    cases are indeed in the linear regime ($n_\sigma\ll 1$) since the
    fermion model~\cite{delvalle09a} gives the same results. The
    system is however almost decoupled due to the large detuning. The
    emission is thus basically that of the bare cavity
    (Lorentzian).}
  \label{fig:MonMay19002338UTC2008}
\end{figure}

In the \emph{non-vanishing} case, of course, the thermal reservoir
gives a different result than unconstrained parameters of our
Eq.~(\ref{eq:SunNov9134217GMT2008}), even in the linear regime (that
is, when $n_\sigma\ll1$, although $n_a$ is not compulsorily also
vanishing). The illustration of this fact was attempted in Fig.~(1a)
of Ref.~\cite{arXiv_ridolfo09a}, although here also the plot is
wrong. With non-vanishing cavity pumping, the linear regime can be
maintained only if the modes are almost uncoupled. This is shown in
our corrected version~(Fig.~\ref{fig:MonMay19002338UTC2008}).

\begin{figure}[tbhp]
  \centering
  \includegraphics[width=.6\linewidth]{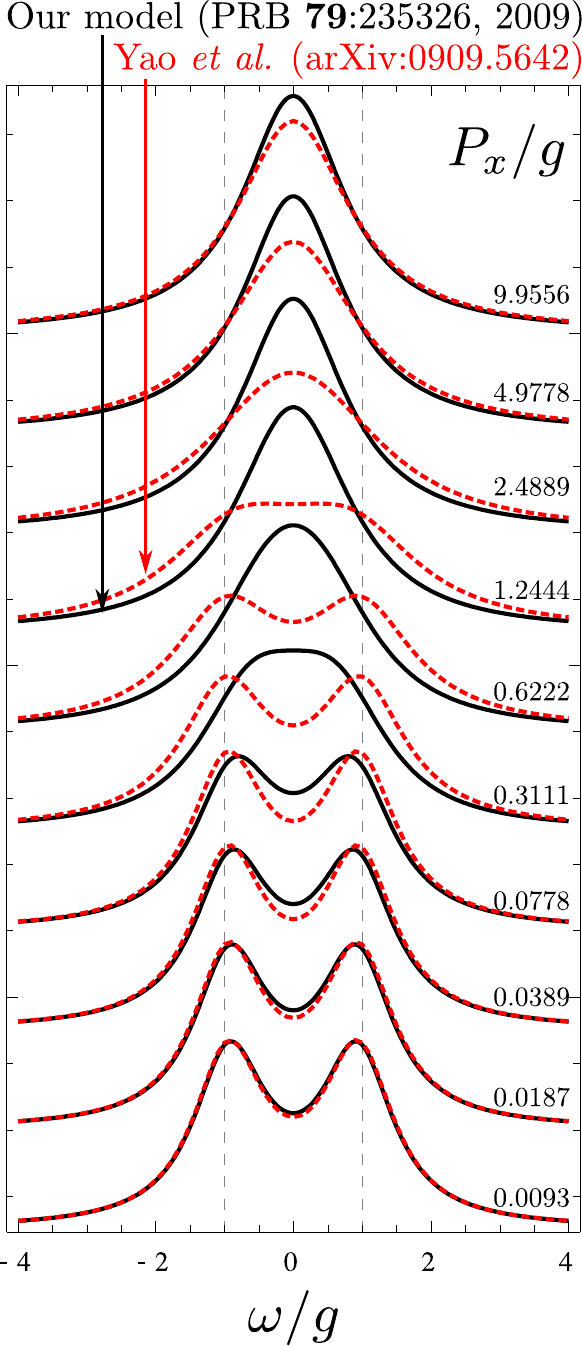}
  \caption{Converged spectra with the full fermion
    model~\cite{delvalle09a} [in solid black] for the choice of
    pumping reservoirs and parameters of Yao \emph{et al.}, along with
    the approximate spectra proposed by these authors [in dashed
    red]. Their approximation is correct only at the smallest values
    of pumping, where it also recovers the boson results of
    Refs.~\cite{laussy08a,laussy09a}.}
  \label{fig:SunOct4214336BST2009}
\end{figure}

\begin{figure}[tbhp]
  \centering
  \includegraphics[width=\linewidth]{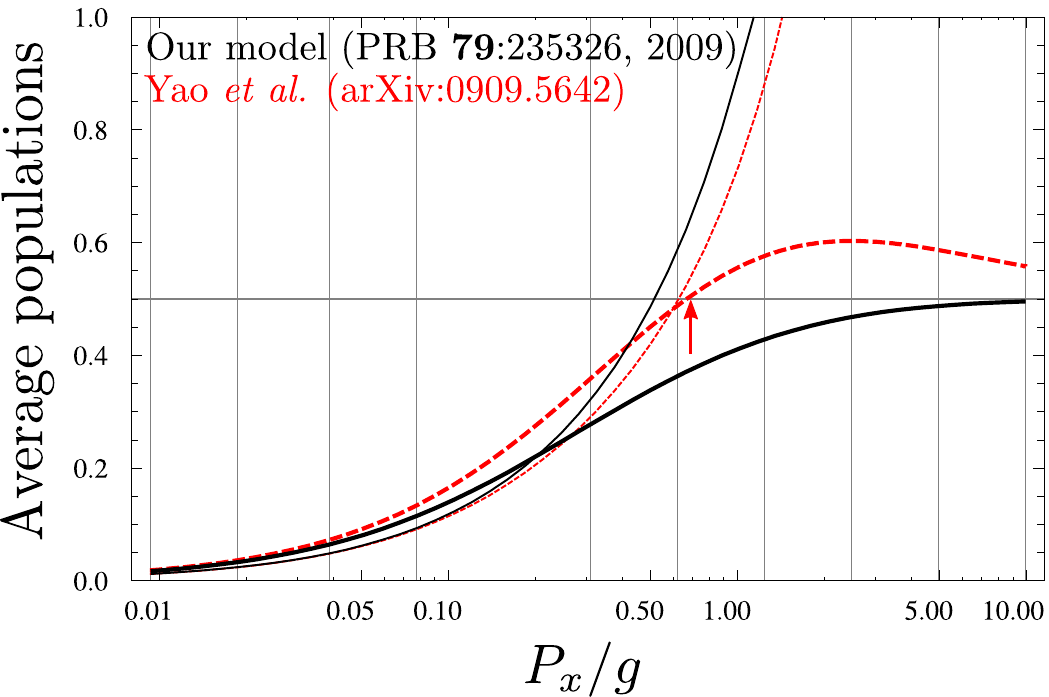}
  \caption{Converged populations [in solid black] for the cavity,
    $\langle\ud{a}a\rangle$ [thin] and the QD,
    $\langle\sigma^+\sigma^-\rangle$ [thick], for the choice of
    pumping reservoirs and parameters of Yao \emph{et al.}, along with
    the approximate values proposed by these authors [in dashed
    red]. Again, their approximation is correct only at the smallest
    values of pumping, where it also recovers the boson results of
    Refs.~\cite{laussy08a,laussy09a}. The breakdown of their
    approximation is further manifest by the inversion of population
    of the QD (indicated by the arrow), which is notoriously
    impossible with thermal reservoirs. Our model shows the expected
    saturation bounded by 1/2.}
  \label{fig:SunOct4215149BST2009}
\end{figure}

We now turn to the approach of Yao \emph{et al.} \cite{arXiv_yao09a}.
They miss a factor $\Gamma_c$ in the second term of the denominator of
their Eq.~(4), although in their case this is a misprint only since
their figures match the formulas that follow from their
approximations. These approximations, however, are incorrect.

As we discussed above, it is straightforward in our approach to
constrain the coefficients, following a given choice of the reservoirs
of excitations. If we choose the thermal baths advocated by Yao
\emph{et al.}, we are able with our approach~\cite{delvalle09a} to
apply the full fermion model with no truncation in the number of
excitations, for the parameters of these authors. We find that their
approximation is a poor one, as seen in
Fig.~\ref{fig:SunOct4214336BST2009}, where we superimpose the exact
(numerical) result, in solid black, to their approximate (analytical)
formula, in dashed red. As could be expected, the agreement is good
only at very low pumping (linear regime) and very high pumping
(uncoupled regime). It is incorrect in the most relevant region of the
transition where the doublet collapses (as has been reported before,
e.g., Fig.~13 of our Ref.~\cite{delvalle09a}).

Their approximation is also basically flawed in that it allows an
inversion of population for the two-level system, although it is
excited by thermal reservoirs, as seen in the magnified version of
their Fig.~3(a), that we reproduce in dashed lines in our
Fig.~\ref{fig:SunOct4215149BST2009}, along with the converged
solutions from our model~\cite{delvalle09a} (with their parameters and
choices of reservoirs). Beyond the poor quantitative agreement when
pumping is non-vanishing, at the point indicated by the arrow and
above, the QD is inverted, which indicates a pathology of their
approximation.

Their implication that cavity pumping is determinant to achieve lasing
is in contradiction with well-known and established facts of the
one-atom laser theory. See for instance our text
Ref.~\cite{delvalle09a} where lasing is achieved without cavity
pumping, thanks to the gain-medium that our general model allows. On
the other hand, thermal reservoirs of Yao \emph{et al.} forbids
lasing, regardless of the magnitude of cavity pumping (note that with
their parameters, very high cavity populations are already achieved,
but they have thermal statistics, with second-order
correlator~$g^{(2)}$ that increases rapidly towards~2 with pumping).

Finally, we want to stress that their Fig.~2(b), that supposedly
represents our model, does not make any meaningful comparison, since,
fitting some data with their model [that we have just shown is wrong,
but even if it was correct], they proceed to plot our \emph{boson}
model with \emph{their} fitting parameters. It is obvious that, the
two formulas being different, the best-fitting parameters for one of
them will yield poor agreement on the same data for the other. Beside,
they should have used our \emph{fermion} model, since they consider a
supposedly two-level emitter in a nonlinear regime. Fitting them
independently, on the one hand, and comparing them on statistical
grounds on the other hand, rather than settling for some aesthetic of
the agreement, is the correct course of action. Fitting with the
nonlinear fermion model is not a trivial task. With E. Cancellieri and
A. Gonzalez-Tudela, we have recently obtained results in this
direction, to be published shortly.

In conclusion, we have shown that our
work~\cite{laussy08a,laussy09a,delvalle09a} is correct and that the
critics addressed against it~\cite{arXiv_ridolfo09a,arXiv_yao09a} are
unsubstantiated on the one hand, and the proposed substitutes are
incorrect on the other hand. These authors do not derive any master
equation. They settle for thermal reservoirs, which derivation is a
standard textbook material. This is a particular case of our work that
they apply incorrectly or beyond its limits of validity. We have
already provided the valid limit for the nonlinear
regime~\cite{delvalle09a}. There remain many open questions in the
field. Some can be settled by statistical analysis of experimental
data with our model, Eq.~(\ref{eq:SunNov9134217GMT2008}), which
correlations between the fitting parameters can teach about underlying
microscopic mechanisms (such as the nature of the bath of excitations,
among other). To this intent, we invite experimentalists to make their
raw data available to
everybody~\footnote{http://sciencecommons.org/about/towards}.


\bibliography{Sci,arXiv,books}\

\end{document}